\documentclass[%
 reprint,
superscriptaddress,
longbibliography,
 amsmath,amssymb,
 aps,
 prb,
]{revtex4-1}

\usepackage{sidecap}
\sidecaptionvpos{figure}{c}

\usepackage{graphicx}
\usepackage{dcolumn}
\usepackage{bm}
\usepackage[
colorlinks=true,
linkcolor=blue,
citecolor=blue,
filecolor=blue,
urlcolor=blue]{hyperref}

\usepackage{amsmath,amsfonts,
	amsgen,amsbsy,amsopn,amstext,amssymb
}
\usepackage{epstopdf}
\usepackage{xcolor}

\usepackage{soul} 

\renewcommand{\i}{i}

\renewcommand{\Pr}{\mathrm{Pr}}

\definecolor{redish}{RGB}{255, 173, 231}
\definecolor{bluish}{RGB}{205, 224, 247}

\begin{document}

\preprint{}

\title{Perfect absorption in mirror-symmetric acoustic metascreens}

\author{V. Romero-Garc\'ia}\email{vicente.romero@univ-lemans.fr}
\affiliation{Laboratoire d'Acoustique de l'Université du Mans (LAUM), UMR 6613, Institut d'Acoustique - Graduate School (IA-GS), CNRS, Le Mans Université, France}
\author{N. Jim\'enez}
\affiliation{Instituto de Instrumentaci\'on para Imagen Molecular, Consejo Superior de Investigaciones Cient\'ificas, Universitat Polit\`ecnica de Val\`encia. Camino de vera s/n 46022 Val\`encia, Spain}
\author{J.-P. Groby}
\affiliation{Laboratoire d'Acoustique de l'Université du Mans (LAUM), UMR 6613, Institut d'Acoustique - Graduate School (IA-GS), CNRS, Le Mans Université, France}
\author{A. Merkel}
\affiliation{Institut Jean Lamour and University of Lorraine in Nancy (France)}
\author{V. Tournat}
\author{G. Theocharis}
\author{O. Richoux}
\author{V. Pagneux}
\affiliation{Laboratoire d'Acoustique de l'Université du Mans (LAUM), UMR 6613, Institut d'Acoustique - Graduate School (IA-GS), CNRS, Le Mans Université, France}

\date{\today}

\begin{abstract}
Mirror-symmetric acoustic metascreens producing perfect absorption independently of the incidence side are theoretically and experimentally reported in this work. The mirror-symmetric resonant building blocks of the metascreen support symmetric and antisymmetric resonances that can be tuned to be at the same frequency (degenerate resonances). The geometry of the building blocks is optimized to critically couple both the symmetric and the antisymmetric resonances at the same frequency allowing perfect absorption of sound from both sides of the metascreen. A hybrid analytical model based on the transfer matrix method and the modal decomposition of the exterior acoustic field is developed to analyze the scattering properties of the metascreen. The resulting geometry is 3D printed and experimentally tested in an impedance tube. Experimental results agree well with the theoretical predictions proving the efficiency of these metascreens for the perfect absorption of sound in the ventilation problems.
\end{abstract}

\pacs{Valid PACS appear here}
\maketitle

\section{Introduction}\label{sec:intro}

Perfect absorption with sub-wavelength materials is a scientific and technological challenge that has received increasing interest during the last years in several branches of wave physics\cite{bliokh2008, Luk2014, piper2014, ma2014, Croenne14, Song2014, Leroy2015}. To meet such challenge, a twofold problem must be solved: the density of states must be increased in the sub-wavelength regime and the impedance of the system must match that of the surrounding medium at the same time. Acoustic metamaterials made of open lossy resonators have been proven to be very good candidates to meet these two conditions\cite{ma2014, Romero2016a, Romero2016b, li2016, Jimenez2016, Jimenez2017b, Lee2019} . They offer the possibility to tune in a controlled manner their inherent losses and, their sub-wavelength character has already been exploited to design perfect absorbers in both the reflection \cite{Jimenez2016, Auregan2018} and the transmission problems\cite{yang2015, Jimenez2017b, Lee2019}.

Previously, perfect absorption has been successfully analyzed using the properties of the scattering matrix of the system\cite{Merkel2015}. In a general manner, perfect absorption can be obtained at the frequencies for which the eigenvalues of the scattering matrix are zero and when the system is excited with the corresponding eigenvectors\cite{Merkel2015}. From the physical point of view, zero eigenvalues of the scattering matrix appear when the critical coupling condition is fulfilled, i.e.,  when the energy leakage of the system is perfectly balanced by its inherent losses\cite{Romero2016b}. In a reflection problem (one port problem), the scattering matrix reduces to the reflection coefficient. The critical coupling conditions have been exploited to design deep sub-wavelength absorbers making use of Helmholtz resonators (HRs)\cite{Romero2016a, Romero2016b, li2016}, membranes, \cite{ma2014, Romero2016a, Duan2015, Wang2018, Auregan2018} coiled-up channels,\cite{ni2014, Zhang16, Yang2017} bubble screens in water \cite{Leroy2015, Lanoy2018}, structured surfaces \cite{Starkey17}, or slow sound metamaterials.\cite{Groby2015, Groby2016, Jimenez2016} In transmission problems (two port system with one-side excitation), the analysis becomes more complicated and depending on the symmetry of the acoustic metamaterial one can deal with two types of problems. If the metamaterial is not mirror-symmetric, the scattering on each side of the structure is different giving rise to different absorption coefficients on the two sides\cite{Fleury15, Auregan17}. At a specific frequency, the critical coupling condition can be fulfilled on one of the sides of the acoustic metamaterial and, as a consequence, single side unidirectional perfect absorption can be obtained\cite{Merkel2015}. This unidirectional perfect absorption has been obtained using HRs to design slow sound type metasurfaces\cite{Jimenez2017b}. If the metamaterial is mirror-symmetric, perfect absorption can be obtained on the two sides of the system but the problem becomes more complicated as a symmetric and an antisymmetric resonance occuring at the same frequency (degenerate resonances) must be activated\cite{piper2014, yang2015}. It is worth noting here that if two resonances in mirror-symmetric systems have the same symmetry, the absorption cannot be perfect. However, quasi-perfect absorption can be obtained using an accumulation point to bring symmetric and antisymmetric resonances as close as possible to each other\cite{Jimenez2017a}. Thus, the main challenge here is the design of the resonant building blocks presenting degenerate resonances.

In this work we design an a metascreen made of mirror-symmetric resonant building blocks possessing degenerate resonances to produce acoustic perfect absorption on the two sides of the metascreen, by the simultaneous critical coupling of both symmetric and antisymmetric resonances. The metascreen consists of a periodic array of slits loaded by different HRs. The resonant mirror-symmetric building block of this metascreen presents both symmetric and antisymmetric resonances allowing for the critical coupling of the transmission problem (two port problem with one-side excitation). The system is analytically studied by a hybrid model combining transfer matrix and modal decomposition methods. The transfer matrix model is used to describe the propagation within each slit while the modal decomposition is used to obtain the propagation in the exterior domain considering the Bloch waves accounting for the possible evanescent coupling between the slits. The two domains are coupled through the continuity conditions between the exterior medium and each slit. A full-wave numerical simulation is used to validate the analytical model. The absorption of the metascreen is optimized for the transmission problem (ventilation), allowing for the perfect absorption in the system from both sides. Finally, the resonant building block is 3D printed and experimentally tested in a square cross-sectional impedance tube, mimicking a metascreen with infinite lateral dimensions. The experimental results agree with the analytical and numerical predictions showing the possibility of designing perfect absorbers for the transmission problem.

\section{Reciprocal and symmetric unidimensional scattering: symmetric and antisymmetric problems}
\label{sec:secII}
In this section we show that the scattering problem by a mirror-symmetric scatterer can be decomposed into two reflection subproblems\cite{Merkel2015, Chesnel18}: ($i$) the symmetric reflection problem, which is produced when the Neumann boundary condition is imposed at the mirror-symmetry plane of the mirror-symmetric resonator, and ($ii$) the antisymmetric reflection problem, which is produced when the Dirichlet boundary condition is imposed at the mirror-symmetry plane of the mirror-symmetric resonator.

\begin{figure}
\includegraphics[width=85mm]{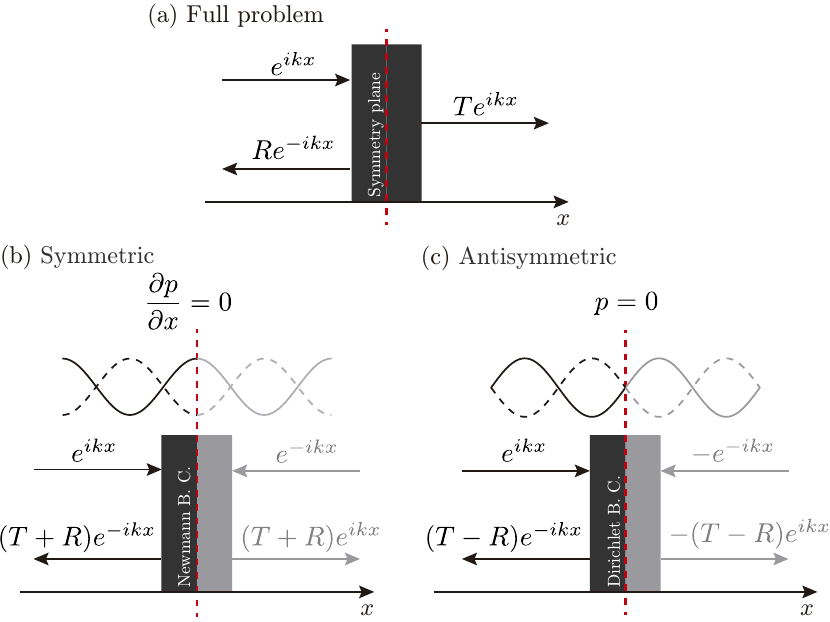}
\caption{(Color online) (a) 1D Transmission problem of a plane wave by a mirror-symmetric resonant scatterer. (b) and (c) show the reflection problem produced when a Neumann and a Dirichlet boundary conditions are respectively considered in the symmetry plane of the symmetric resonant scatterer (red dashed lines in (b) and (c)). Continuous line represents the incident pressure wave and dashed line represents the reflected wave due to the boundary condition. In gray we represents the image generated by the presence of the boundary condition.}
\label{fig:fig1}
\end{figure}

A reciprocal and symmetric unidimensional (1D) scattering problem is characterized by a symmetric scattering matrix,
\begin{eqnarray}
S(\omega)=\left(\begin{tabular}{cc}
$T$ & $R$\\
$R$ & $T$
\end{tabular}\right),
\end{eqnarray}
where $R$ and $T$ are the complex frequency dependent reflection and transmission coefficients of the system respectively. The absorption coefficient can thus be defined as $\alpha=1-|R|^2-|T|^2$. The eigenvalues of the scattering matrix are given by $\lambda_{\pm}=T\pm R$, while the eigenvectors are $\vec{v}_{\pm}=(1,\pm1)$. Figure~\ref{fig:fig1}(a) represents the scattering problem when a symmetric system (symmetry with respect to the median plane [white dashed line in Fig.~\ref{fig:fig1}(a)]) is excited by a plane wave from the left with the harmonic time convention $e^{-i\omega t}$. First, we consider that a Neumann boundary condition $\partial p/\partial x=0$ is imposed at the symmetry plane of the system, $p$ representing the pressure wave which is the solution of the Helmholtz equation (see Fig.~\ref{fig:fig1}(b)). The Neumann boundary condition imposes a perfect mirror symmetry creating a virtual image of the full problem on the right side. Therefore, the reflection coefficient of this particular subproblem, $R_s$, can be calculated from the reflection and transmission coefficients of the original problem, $R_s=T+R$. The subindex $s$ means symmetric since the Neumann boundary condition selects only modes that are symmetric with respect to the symmetry plane where the condition is imposed. Second, we consider that a Dirichlet boundary condition, $p=0$, is imposed at the mirror-symmetry plane of the system. This plane acts as a mirror creating a virtual image on the right side with a $\pi$ phase shift of the full problem. The reflection coefficient of this particular subproblem, $R_a$, can be calculated from the reflection and transmission coefficients of the original one, $R_a=T-R$. The subindex $a$ means antisymmetric since the Dirichlet boundary condition imposes only antisysymmetric modes with respect to the symmetry plane. Note that the two eigenvalues of the scattering matrix of the original problem, $\lambda_+$ and $\lambda_-$, correspond to the reflection coefficients of the reflection subproblems with Neumann and Dirichlet conditions, respectively. In other words, the original transmission problem can be decomposed in the symmetric reflection problem, $\lambda_+=R_s$, and the antisymmetric reflection problem, $\lambda_-=R_a$. Thus, the original scattering coefficients read as
\begin{eqnarray}
R=\frac{R_s-R_a}{2},\;\;\;\;
T=\frac{R_s+R_a}{2},\;\;\;\;
\alpha=\frac{\alpha_s+\alpha_a}{2},
\end{eqnarray}
where $\alpha_s=1-|R_s|^2$ and $\alpha_a=1-|R_a|^2$ are the absorption coefficients of the reflection subproblems with Neumann and Dirichlet boundary conditions respectively. 

\begin{figure*}
\includegraphics[width=170mm]{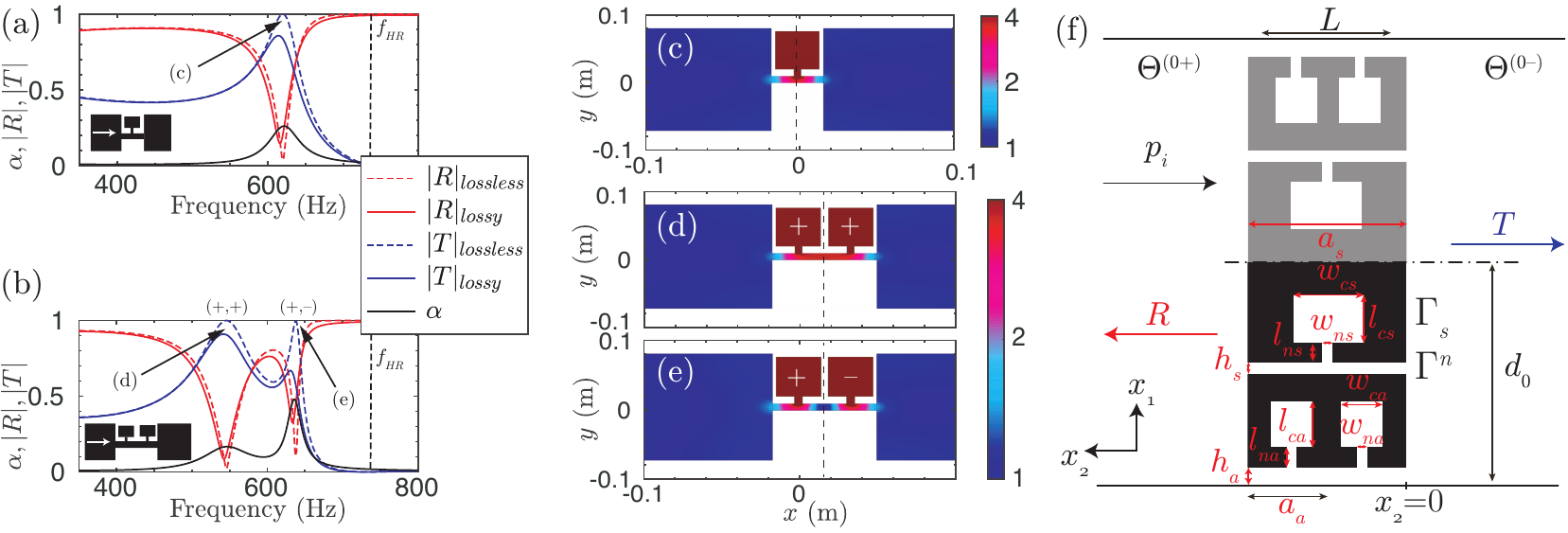}
\caption{(Color online) (a) and (b) show the lossless (dashed lines) and lossy (continuous lines) scattering coefficients for the case of a slit loaded by one and two identical resonators respectively. (c) Absolute value pressure field for the frequency of the first resonance of the slit loaded with one resonator. Vertical dashed line shows the symmetry plane of the metascreen. (d) and (e) represent the absolute value pressure field for the two first resonances of the slit loaded with two identical resonators. Vertical dashed line shows the symmetry plane of the metascreen. (f) Scheme of the mirror-symmetric building block used in this work with the definitions of the geometrical parameters.}
\label{fig:fig2}
\end{figure*}

\section{Mirror-symmetric resonators with degenerate resonances for the ventilation problem}
Perfect absorption in 1D reciprocal problems with one-side excitation appears when the two eigenvalues of the scattering matrix simultaneously vanish at the same frequency\cite{Merkel2015, Jimenez2017b}. In other words, perfect absorption in mirror-symmetric 1D reciprocal problems is achieved when the symmetric and the antisymmetric reflection problems are critically coupled simultaneously at the same frequency. In this case, $\alpha_s=\alpha_a=1$ and thus $\alpha=1$. This means that we need degenerate resonances to perfectly absorb waves, i.e., resonators with both symmetric and antisymmetric resonances at the same frequency. Note that, if either symmetric or antisymmetric resonances are excited, only half of the problem can be critically coupled, thus the maximal absorption that could be reached is $\alpha=0.5$.

In this section we use systems made of slits loaded by different HRs embedded in a waveguide of different cross section. First, we analyze the propagation in a slit loaded with a single resonator as shown in the inset of Fig.~\ref{fig:fig2}(a). We choose a resonator with a resonance frequency at $f_{H\!R}=735$ Hz. The scattering coefficients of the system are shown in Fig.~\ref{fig:fig2}(a). The first mode of the slit, i.e. a Fabry-Perot resonance, is shifted at a frequency lower that $f_{HR}$ because of the strong dispersion introduced by the loading resonator. This mode translates in the form of a transmission peak. Figure~\ref{fig:fig2}(c) shows the absolute value of the pressure field at this particular frequency, showing the first resonance mode of the slit. It is worth noting here that this resonance mode is symmetric with respect to the symmetry plane of the system. Second, we analyze the propagation through a slit loaded by two HRs as shown in the inset of the Fig.~\ref{fig:fig2}(b). In this case the scattering coefficients are plotted in Fig.~\ref{fig:fig2}(b), showing two transmission peaks corresponding to the first two Fabry-Perot resonances of the slit. The two first resonances of the slit are clearly visible in Figs. \ref{fig:fig2}(d,e) that depict the absolute value of the pressure field at these particular frequencies. The first mode is symmetric while the second mode is antisymmetric with respect to the symmetry plane of the system (shown by the dashed line in Figs. \ref{fig:fig2}(c-e)).

From the previous discussion in Section \ref{sec:secII} we can conclude that resonators with symmetric and antisymmetric resonances can be designed with the combination of the two discussed subproblems, a periodic arrangement of two slits in the $x_1$ direction loaded respectively with one and two HRs. As a consequence, the metascreen  design, shown in Fig.~\ref{fig:fig2}(f), is a combination of slits loaded by one and two resonators. The slit loaded by a single resonator only supports symmetric modes, and  will thus be named symmetric slit. The slit loaded by two resonators supports both symmetric and antisymmetric modes. It will be mainly used to tune the antisymmetric resonance and will thus be named antisymmetric slit. The geometric parameters of the metascreen are defined in Fig.~\ref{fig:fig2}(f). The subindex $s$ ($a$) represents the parameters for the symmetric (antisymmetric) slit. The HRs are made by combining two different elements playing the role of the neck and the cavity. The problem consists in critically coupling the symmetric and the antisymmetric slits at the same frequency by optimizing the geometry, which controls the energy leakage and the inherent losses (viscothermal losses).

At this stage we have to discuss the unit cell geometry shown in Fig.~\ref{fig:fig2}(f). For the experiments, we only have to consider half of it [black areas in Fig.~\ref{fig:fig2}(f)] as we are using a square impedance tube of side $d_0$. In fact, the walls of the impedance tube act as mirrors generating the virtual images (grey areas in Fig.~\ref{fig:fig2}(f)) with respect to the position of the rigid wall in the tube, i.e., the dot-dashed line in Fig.~\ref{fig:fig2}(f). 

\section{Modeling}
The theoretical approach considered in this work is a hybrid model combining modal decomposition and transfer matrix methods. The modal decomposition of the acoustic field is used at the exterior domains in order to account for the Bloch modes excited at the interfaces of the metascreen allowing possibly the evanescent coupling between the different building blocks. Wave propagation in each slit is modeled by the transfer matrix method as widely done previously \cite{Jimenez2016, Jimenez2017b}.

The acoustic field in the exterior domains ($\Theta^{(0+)}$ and $\Theta^{(0-)}$ as defined in Fig.~\ref{fig:fig2}(f)), when the system is excited by a plane wave $\vec{k}^0=(k_1^i,k_2^i)$ can be written as
\begin{align}
p^{(0+)}=&\sum_q\left[e^{-\imath k_{2q}^{0}(x_2-L)}\delta_q\right.\nonumber\\
&\quad\quad\quad+\left.R_qe^{\imath k_{2q}^{0}(x_2-L)}\right]e^{\imath k_{1q}^{0}x_1}\;\;\textrm{in}\;\;\Theta^{(0+)},
\\
p^{(0-)}=&\sum_q T_q e^{\imath k_{1q}^{0}x_1-\imath k_{2q}^{0}x_2}\;\;\textrm{in}\;\;\Theta^{(0-)},
\end{align}
where the subindex $q$ indicates the order of the Bloch wave, with $k_{1q}^0=2q\pi/2d_0+k_1^i$ and  $k_{2q}^0=\sqrt{(|\vec{k}^0|)^2-(k_{1q}^0)^2}$, with $\textrm{Re}\left( k_{2q}^0\right)\geq 0$, $R_q$ and $T_q$ are the reflection and transmission coefficients of the $q$-th Bloch wave, $\delta_q$ is the Kronecker's delta and $L$ is the thickness of the acoustic metascreen.

Wave propagation in each slit is modeled by the transfer matrix method (TMM), in which the HRs are considered as 1D point scatterers\cite{Jimenez2016, Jimenez2017b}. Matrices describing the propagation in the slit, to the side branch resonators and to the section mismatch are  assembled giving  the total transfer  matrix\cite{Jimenez2016, Jimenez2017b}. The  TMM  assumes  the  plane-wave propagation in each element, which is valid in the low frequency regime. The radiation corrections are included in the impedances of the resonators to mimic the effect of the higher order modes. Therefore, the wave propagation through the slit $n$ can be modeled as
\begin{eqnarray}\label{eq:Ttotal}
\left[\begin{tabular}{c}
$p^{(n)}$\\
$v_2^{(n)}$
\end{tabular}\right]_{x_2=L} & = & {\bf{T}}^{(n)} \left[\begin{tabular}{c}
$p^{(n)}$\\
$v_2^{(n)}$
\end{tabular}\right]_{x_2=0} \nonumber\\
&=&
\left[\begin{tabular}{cc}
$T_{11}^{(n)}$ & $T_{12}^{(n)}$\\
$T_{21}^{(n)}$ & $T_{22}^{(n)}$
\end{tabular}\right]
\left[\begin{tabular}{c}
$p^{(n)}$\\
$v_2^{(n)}$
\end{tabular}\right]_{x_2=0},
\end{eqnarray}
where  ${\bf{T}}^{(n)}$ is given by the product of the transfer matrices of the $N$ elements in the $n$-th slit as
\begin{equation}\label{eq:totalmatrix}
{\bf{T}}^{(n)} =  \prod_{j=1}^{N} {\bf T}_j^{(n)} \,.
\end{equation}
The transfer matrices ${\bf T}_j^{(n)}$, is calculated accordingly to the nature of the element and are given in detail in the Supplementary material.

\begin{figure}
\includegraphics[width=85mm]{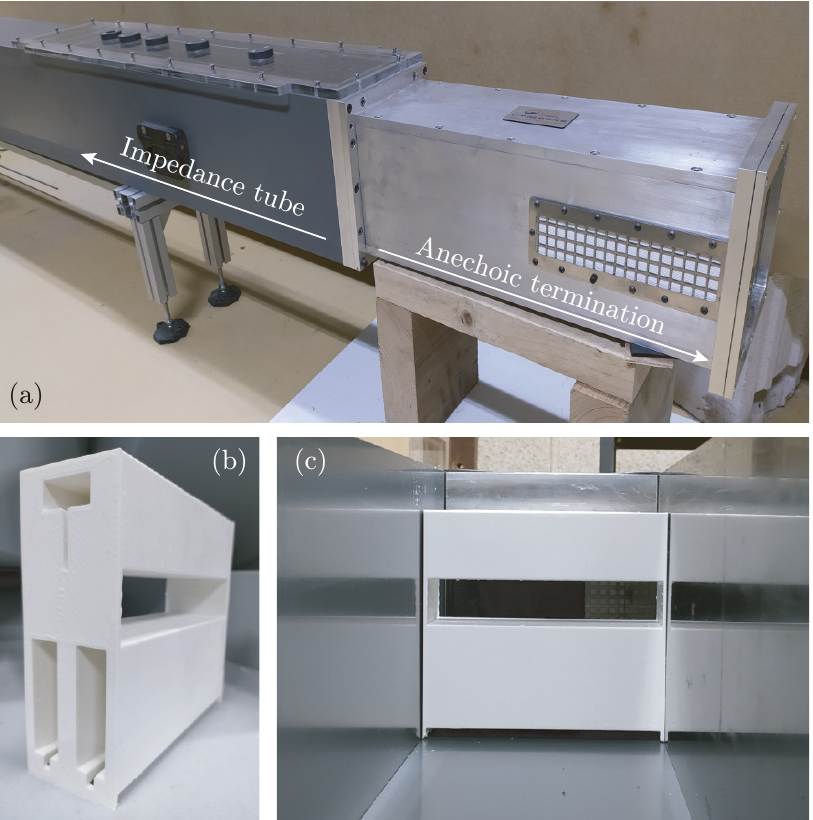}
\caption{(Color online) (a)  Picture of the impedance tube used for the acoustic measurements with the anechoic termination. (b) Picture of the 3D printed symmetric resonant scatterer. (c) Picture of the symmetric resonant scatterer inside the impedance tube.}
\label{fig:fig3}
\end{figure}

\begin{figure*}
\includegraphics[width=170mm]{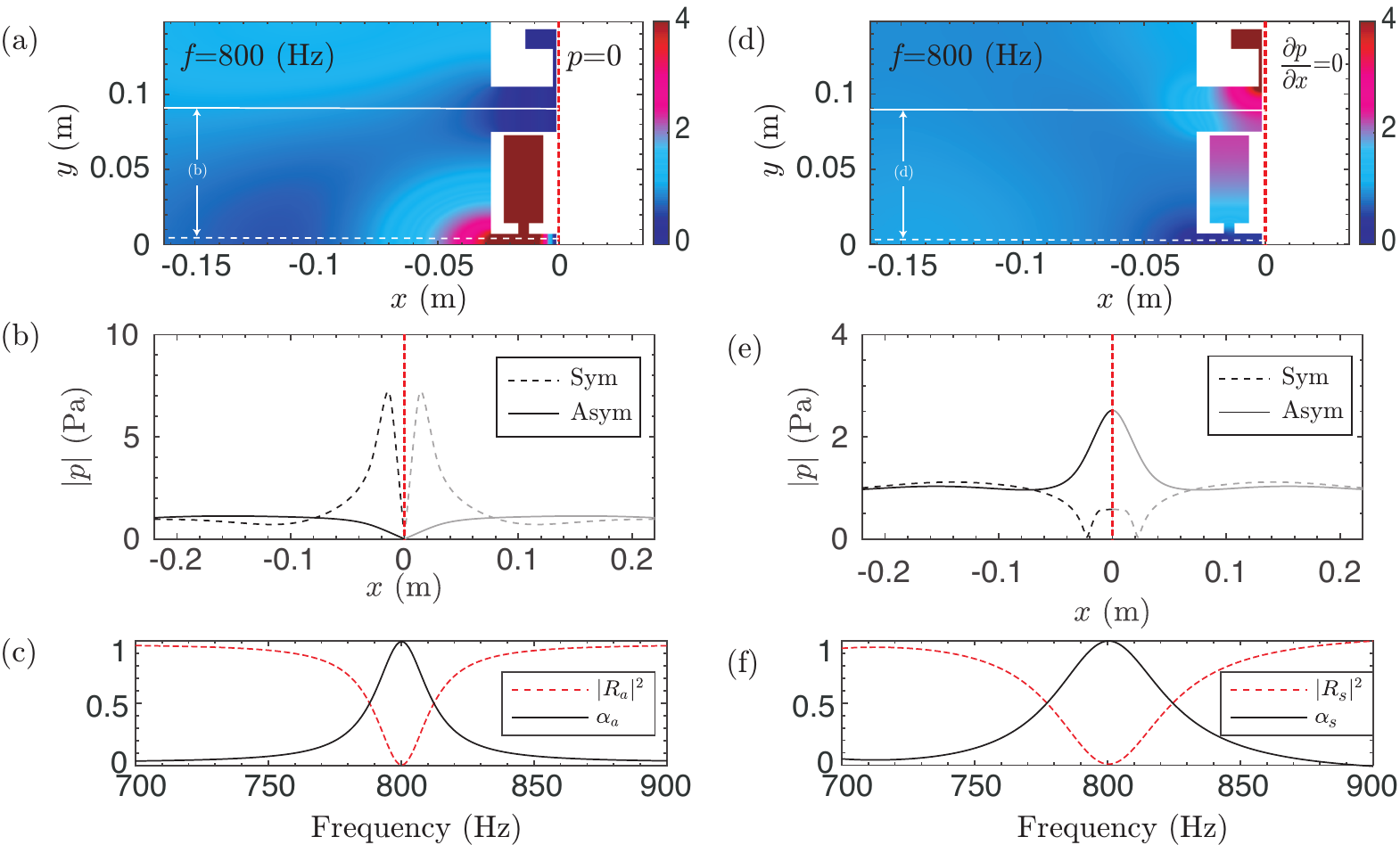}
\caption{(Color online) Critical coupling of the symmetric and antisymmetric problem. (a) Represents the absolute value distribution of the acoustic field at the perfect absorption of the antisymmetric problem. (b) Shows the absolute value, $|p|$, along the profiles shown in (a) considering the field in each slit for the antisymmetric problem. (c) Shows the reflection and the absorption coefficient of the antisymmetric problem. (d) Represents the absolute value distribution of the acoustic field at the perfect absorption of the symmetric problem. (e) Shows the absolute value, $|p|$, along the profiles shown in (c) considering the field in each slit for the symmetric problem. (f) Shows the reflection and the absorption coefficient of the symmetric problem.}
\label{fig:fig4}
\end{figure*}

Once the acoustic fields inside each domain of the problem are defined, the continuity boundary conditions must be applied. These conditions read as
\begin{eqnarray}
p^{(0+)}&=&p^n\;\;\textrm{on}\;\;\Gamma^{n}\;\;\textrm{at}\;\;x_2=L,\\
v_2^{(0+)}&=&v_2^n\;\;\textrm{on}\;\;\Gamma^{n}\;\;\textrm{at}\;\;x_2=L,\\
v_2^{(0+)}&=&0\;\;\textrm{on}\;\;\Gamma_s\;\;\textrm{at}\;\;x_2=L,\\
p^{(0-)}&=&p^n\;\;\textrm{on}\;\;\Gamma^n\;\;\textrm{at}\;\;x_2=0,\\
v_2^{(0-)}&=&v_2^n\;\;\textrm{on}\;\;\Gamma^n\;\;\textrm{at}\;\;x_2=0,\\
v_2^{(0-)}&=&0\;\;\textrm{on}\;\;\Gamma_s\;\;\textrm{at}\;\;x_2=0,
\end{eqnarray}
where $\Gamma_s$ represents the rigid interfaces on the two faces of the metascreen, and $\Gamma^n$ is the interface between the $n$-th slit and the surrounding media $\Theta^{(0^+)}$ and $\Theta^{(0^-)}$. These conditions lead to a system of equations after projection on the adequate orthogonal modes. The system is solved for the scattering coefficients of the system. Details about these calculations are given in the Supplementary material.

In order to validate this model, we have used a finite element method (FEM) simulation using COMSOL Multiphysics software to solve the scattering problem. A plane wave impinges the system and the complete geometry is considered using the radiation conditions that simulate the Sommerfeld conditions at the limits of the numerical domain.

In the model and in the FEM simulations the viscothermal losses are introduced via effective parameters in every narrow region, i.e., the slits that compose the main waveguide and the resonators\cite{stinson1991}. The effective bulk modulus and mass density of a slit of width $h$ are given by
\begin{align}
\rho_s&={\rho _0}\left[1-\frac{\tanh \frac{h}{2}{G_\rho}}{\frac{h}{2}{G_\rho}}\right]^{-1} \,,\label{eq:rho_s:slit}\\
\kappa_s&=\kappa_0\left[1+(\gamma-1)\frac{\tanh \frac{h}{2}{G_\kappa} }{\frac{h}{2}{G_\rho}}\right]^{-1}\,,\label{eq:K_s:slit}
\end{align}
where $h$ is the width of the slit, $G_\rho=\sqrt{{i\omega\rho_0}/{\eta}}$ and $G_\kappa=\sqrt{i\omega\mathrm{Pr}\rho_0/{\eta}}$, 
$\gamma$ is the specific heat ratio of the fluid, $P_0$ is the atmospheric pressure, $\Pr$ is the Prandtl number, $\eta$ the dynamic viscosity, $\rho_0$ the fluid density, $\kappa_0={\gamma P_0}$ the fluid bulk modulus, in our case air.

\begin{table}
\begin{tabular}{|c|c|c|c|c|c|} 
 \hline
 $a_s$ & $h_s$ & $w_{cs}$ & $l_{cs}$ & $w_{ns}$ & $l_{ns}$ \\
(mm) & (mm) & (mm) & (mm) & (mm) & (mm) \\
 \hline
53.6 & 29.8 & 25.45 & 12.9 & 2.4 & 25.1\\
 \hline\hline
$a_a$ & $h_a$ & $w_{ca}$ & $l_{ca}$ & $w_{na}$ & $l_{na}$ \\
(mm) & (mm) & (mm) & (mm) & (mm) & (mm) \\
 \hline
26.8 & 6.6 & 16.05 & 58 & 4.2 & 7.1\\
 \hline\hline

 \hline
\end{tabular}
\caption{Geometrical parameters for the symmetric and antisymmetric resonators of the metascreen for perfect absorption at 800 Hz.}
\label{tab:tab1}
\end{table}

\begin{figure*}
\includegraphics[width=170mm]{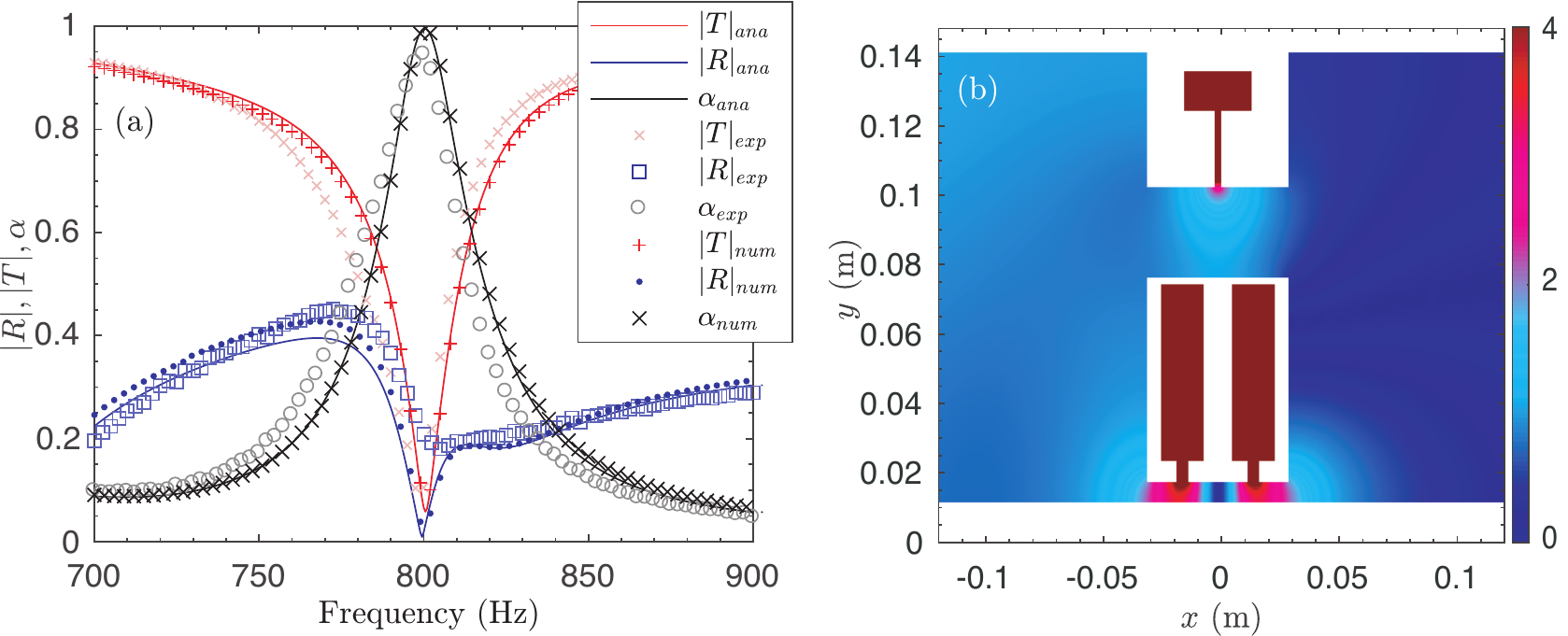}
\caption{(Color online) (a) Analytical, numerical and experimental scattering coefficients of the full problem. (b) Represents the absolute value distribution of the acoustic field at the perfect absorption of the full scattering problem.}
\label{fig:fig5}
\end{figure*}

\section{Experimental set-up}
The reflection and transmission coefficients of a symmetric scatterer can be recovered from four-microphone impedance tube measurements. Figure~\ref{fig:fig3}(a) shows the picture of the square cross-sectional impedance tube used for the characterization of the metascreen designed in this work. The cross section of the impedance tube has a surface of $150\times150$ mm$^2$, which allows the acoustic characterization of materials in the range of frequencies between 50 Hz and 1200 Hz. The impedance tube is terminated by an anechoic termination as shown in Fig.~\ref{fig:fig3}(a). The amplitude of the acoustic source is low enough to neglect the nonlinear behavior of the HRs.

\section{Results}
The model is used to calculate the absorption coefficient of the designed metascreen. This model is coupled with an optimization method (sequential quadratic programming (SQP) method) that varies the geometry of the system providing structures with perfect absorption. The thickness of the metascreen, $L$ and the target frequency at which perfect absorption is desired, are fixed in such a way that the system presents sub-wavelength dimensions. The geometric parameters for a perfect absorption at 800 Hz with $\lambda/L\simeq 8$ are shown in Tab.~\ref{tab:tab1} (where $\lambda$ is the wavelength of the wave at the target frequency in air). Note that these fixed quantities impose $a_s=2a_a=L$. Note also that the distance $d_0=148$ mm for the resonant building block is chosen slightly smaller than the side of the square impedance tube to allow a perfect fit of the sample using vacuum grease at the boundaries.

In this section we explain from a physical point of view the optimization process steps to obtain the perfect absorption in the transmission problem. 

\subsection{Critical coupling of the symmetric and the antisymmetric problems}
The antisymmetric slit supports two resonant modes, one symmetric and one antisymmetric, while the symmetric slit supports only one symmetric mode. Thus, due to the geometry considered in this work only one antisymmetric resonant mode can be excited in the antisymmetric slit. Therefore, we start the discussion with the antisymmetric subproblem. Consider a Dirichlet condition in the mirror-symmetry plane of the metascreen. In the symmetric slit only symmetric modes can be excited. Therefore, no modes will be excited inside this slit when imposing a Dirichlet boundary condition. However, in the antisymmetric slit, only the antisymmetric mode will be excited. Thus, the antisymmetric subproblem can
be critically coupled by optimizing only the geometry of the antisymmetric slit. Figures~\ref{fig:fig4}(a,b) show the acoustic field of the antisymmetric reflection problem at the frequency with perfect absorption. Only the antisymmetric slit is excited. The reflection and absorption coefficients of this problem are shown in Fig.~\ref{fig:fig4}(c). At the target frequency, 800 Hz, perfect absorption is obtained. 

Once the antisymmetric subproblem is critically coupled, we have to consider the symmetric one by imposing a Neumann boundary condition at the mirror-symmetry plane of the metascreen. In that case, symmetric modes exists in both slits. However, as the antisymmetric problem has been critically coupled by optimizing the geometry of the antisymmetric slit, now we only use the geometry of the symmetric slit to critically couple the symmetric subproblem. Figures~\ref{fig:fig4}(d,e) show the acoustic field of the symmetric reflection problem at the frequency with perfect absorption. In this subproblem, the two resonators are excited. The reflection and the absorption coefficients of this subproblem are shown in Fig.~\ref{fig:fig4}(f).

\subsection{Perfect symmetric absorption}

In the previous section we have critically coupled the symmetric and the antisymmetric reflection subproblems. This means that the two eigenvalues of the scattering matrix vanish at the same frequency for the full transmission problem with the optimized geometry. Therefore, the full problem is critically coupled and perfect absorption is expected. The geometry obtained from the optimization problem for the complete resonator is used to 3D print the resonator as shown in Fig.~\ref{fig:fig3}(b). The resonator is mounted in the square cross-sectional impedance tube (Fig.~\ref{fig:fig3}(c)) and the scattering parameters of the full problem are measured. 

Figure~\ref{fig:fig5}(a) shows the scattering coefficients of the full transmission problem. The agreement between the model, the numerical simulations and the experimental results is very good. Note here that the presence of the Bloch waves is crucial in this kind of systems: if only plane waves had been accounted for in the system, the model would not have reproduced the whole wave process because the evanescent coupling would have been neglected. When the Bloch waves are considered, a perfect absorption peak is observed at 800 Hz as analytically, numerically and experimentally observed. Figure~\ref{fig:fig5}(b) shows the acoustic field of the full problem at the perfect absorption peak. It is worth noting here that perfect absorption is very sensible to the geometry of the resonators, and this would explain the slight discrepancies between the analytical or numerical predictions and the experimental results. The symmetric slit exhibits the symmetric Fabry-Perot mode while the antisymmetric slit exhibits the antisymmetric Fabry-Perot mode. Both modes are excited at the same frequency, i.e., the structure presents a degenerate resonance.

\section{Conclusions}

Two subsystems with a degenerate resonance (symmetric for one and antisymmetric for the second) have been designed as building blocks of an acoustic metascreen presenting perfect absorption independently of the incidence side. The resonant scattering produced by the metascreen is analytically studied by a hybrid model mixing the transfer matrix method and the modal decomposition accounting for the Bloch waves in order to consider the possible coupling between the slits. Note that this coupling is particularly important for the present topology. If no higher order Bloch waves had ben accounted for, i.e., if only the plane wave had been accounted for outside the metascreen, the analytical model would not have been in agreement with neither the full-wave numerical simulation nor the experimental results. The structure is experimentally tested in a square impedance tube showing good agreement with the analytical and numerical predictions. Two facts should be highlighted. First, the metascreen is made of acoustically rigid materials without any element vibrating. The life duration of the present metascreen is thus expected to be longer than that of metascreens composed of membranes or plates. Second, the height of the slits of the proposed metascreen are large enough for the system to be used in the ventilation problems where air flow is of imporatnce. The use of other types of resonators as 3D HRs will allow to reach thiner structures with deeper sub-wavelength dimensions. The Physics behind this work will motivate the development of broadband and perfect absorption for ventilation problems using cascade processes similar to the case of the rainbow trapping for the unidirectional absorption\cite{Jimenez2017b}.

\begin{acknowledgments}	
The authors gratefully acknowledge the ANR-RGC METARoom (ANR-18-CE08-0021) project, the project HYPERMETA funded under the program \'Etoiles Montantes of the R\'egion Pays de la Loire. This article is based upon work from COST Action DENORMS CA15125, supported by COST (European Cooperation in Science and Technology). N.J. acknowledges financial support from the Spanish Ministry of Science, Innovation and Universities (MICINN) through grant ``Juan de la Cierva-Incorporaci\'on'' (IJC2018-037897-I).
\end{acknowledgments} 


%

\end{document}